\documentstyle[bbm,amsfonts,epsfig,12pt,here]{article}

\newcommand {\slsh} [1] {\not{\hbox{\kern-2pt${#1}$}}}

\newcommand {\beqn}{\begin{eqnarray}}
\newcommand {\eeqn} {\end{eqnarray}}

\newcommand {\beq} {\begin{equation}}
\newcommand {\eeq} {\end{equation}}
\newcommand {\ber}{\begin{eqnarray*}}
\newcommand {\eer} {\end{eqnarray*}}
\newcommand {\bea}{\begin{eqnarray}}
\newcommand {\eea} {\end{eqnarray}}
\newcommand{\Nfour} {${\cal N}=4\ $}

\newcommand{\None}{${\cal N}=1\ $}
\newcommand{\ztwo}{${Z}_2\ $}

\newcommand {\eqref} [1] {(\ref {#1})}

\begin{document}
\begin{titlepage}
\begin{flushright}{FTPI-MINN-05/08,
UMN-TH-2350/05\\
SWAT/05/429\\
ITEP-TH-29/05\\
DRAFT
}
\end{flushright}

\vskip 0.4cm

\centerline{{\Large \bf Spontaneous \boldmath{\ztwo} Symmetry Breaking
 in the Orbifold}}

\vspace{2mm}

\centerline{{\Large \bf  Daughter of \boldmath{\None} $\!\!\!\!$
Super-Yang--Mills Theory,}}

\vspace{2mm}

\centerline{{\Large \bf Fractional Domain Walls and Vacuum Structure}}

\vskip 0.5cm
\centerline{\large A. Armoni $^{a}$, A. Gorsky $^{b,c}$,
and M. Shifman $^{b}$}

\vskip 0.3cm

\centerline{$^a$\em Department of Physics, University of Wales Swansea,}
\centerline{\it Singleton Park, Swansea, SA2 8PP, UK}
\vspace{2mm}
\centerline{$^b$\em William I. Fine Theoretical Physics Institute, University of Minnesota,}
\centerline{\it Minneapolis, MN 55455, USA}
\vspace{2mm}
\centerline{$^c$ \it  Institute of Theoretical and Experimental Physics, Moscow
117259, Russia}

\vskip1.3cm

\begin{abstract}

We discuss the fate of the \ztwo symmetry and the vacuum structure
in an SU$(N)\times$SU($N$) gauge theory with one
bifundamental Dirac fermion. This theory can be obtained from
SU($2N$) supersymmetric Yang--Mills (SYM)
theory by virtue of \ztwo orbifolding. We analyze  dynamics of domain
walls and argue that the $Z_2$ symmetry is spontaneously broken.
Since unbroken $Z_2$ is a necessary condition
for nonperturbative planar equivalence we conclude that the orbifold
daughter is nonperturbatively nonequivalent to its supersymmetric parent.
{\em En route},
our investigation reveals the existence of {\em fractional} domain walls,
similar to fractional D-branes of string theory on orbifolds.
We conjecture on the fate of these domain walls in the true solution
of the $Z_2$-broken orbifold theory.
We also comment on  relation with nonsupersymmetric string
theories and closed-string tachyon condensation.

\end{abstract}

\thispagestyle{empty}
\end{titlepage}
\tableofcontents

\newpage

\section{Introduction}
\label{intro}

Recently, a considerable progress has been achieved
\cite{AAMSGV,ASV-one,ASV-two}
in understanding of nonsupersym\-metric Yang--Mills
theories which can be obtained from supersymmetric
gluodynamics by orbifolding or orientifolding,
following the original discovery of planar equivalence
\cite{kachru,Lawrence:1998ja,K,KK,Kakushadze,schmaltz,armoni-kol}.
While establishing  perturbative planar equivalence is quite straightforward,
the issue of nonperturbative  equivalence
of the orbi/orientifold daughters to the parent theory ---
supersymmetric  gluodynamics --- is more complicated.
The question of  nonperturbative equivalence between
supersymmetric (SUSY) and non-SUSY theories
was raised by Strassler \cite{strassler} who formulated
a {\em nonperturbative orbifold conjecture} (NPO).
Shortly after Strassler's work, arguments were given   \cite{GS,dt} that
in the orbifold daughter
planar equivalence  fails at the
nonperturbative level. In particular, Tong
showed that when the orbifold theory is compactified on a
spatial circle, the SYM-inherited vacuum is not the
genuine vacuum of the theory \cite{dt}.
It was discovered, however,
that the orientifold daughter is more robust and withstands
the passage to the nonperturbative level \cite{AAMSGV,ASV-one,ASV-two}.

A refined proof of the  nonperturbative equivalence
of the orientifold daughter was worked out in Ref.~\cite{ASV-two}.
Here we carry out a similar analysis for the orbifold
theory. This analysis will show, in a very transparent manner,
that the necessary condition for the nonperturbative equivalence
to hold in the  orbifold case is that the \ztwo symmetry of the ($Z_2$) orbifold Lagrangian is not spontaneously broken.
The same conclusion was reached in \cite{Yaffe}.

As well-known \cite{klebanov-tseytlin,klebanov},
string theory prompts us that,
for the orbifold daughter of \Nfour SYM theory, the \ztwo symmetry
is spontaneously
broken above a critical value of the 't Hooft coupling. The orbifold
field theory under consideration can be described by
a brane configuration of type-0 string theory \cite{armoni-kol} (see Sect.~\ref{sectype0}). Type-0 strings contain a closed-string
tachyon mode in the twisted sector. The tachyon couples to
the  ``twisted" field  \cite{klebanov}
\beq
T \equiv \left( {\rm Tr} \, F_{ e} ^2 - {\rm Tr} \, F_{  m}^2\right)
\label{iman}
\eeq
of the SU$_e(N)\times$SU$_m(N)$ gauge theory.\footnote{Here and below the normalization of  traces is such that
$$
{\rm Tr}\, F\tilde F = \sum_{a=1}^{4N^2}F_{\mu\nu}^a\, \tilde  F^{\mu\nu\,\,  a}\,,
\qquad {\rm Tr}\,( F\tilde F)_e
= \sum_{a=1}^{N^2}\left( F_{\mu\nu}^a\, \tilde  F^{\mu\nu\,\,  a}\right)_e \,,
$$
and so on.}
The subscripts $e$ and $m$ refer to ``electric" and ``magnetic",  respectively. The words electric and magnetic, borrowed from the string theory
terminology, are used here just to distinguish between the two SU$(N)$'s of the gauge group SU$(N)\times$SU($N$).

The prediction of string theory \cite{klebanov} is that the perturbative vacuum at $\langle T \rangle=0$ is unstable. In the {\em bona fide} vacua a condensate of the form
\beq
\left\langle {\rm Tr} \, F_e ^2 - {\rm Tr} \, F_m ^2 \right\rangle = \pm \Lambda ^4
\label{cotf}
\eeq
must develop.

Our task is to explore this phenomenon
within field theory {\em per se}, with no (or almost no) reference to string theory.
The SU$_e(N)\times$SU$_m(N)$ gauge theory with a Dirac bifundamental field is
very  interesting by itself, with no reference to orbifolding.
If we could {\em prove} that \ztwo is spontaneously broken,
using field-theoretic methods, this would be a tantalizing development. Below we will
present arguments that such spontaneous symmetry breaking does take place,
which, although convincing, stop short of being a full proof.

First, we generalize the analysis of Ref.~\cite{ASV-two} to demonstrate
that NPO does require unbroken $Z_2$.
Then we
proceed to  arguments based on consideration of domain wall dynamics
to show that the domain wall of the parent SYM theory, upon
orbifolding, becomes unstable and splits into two walls:
one ``electric" and one ``magnetic." As will be explained
below, this splitting is a  signal of the spontaneous breaking of
the \ztwo symmetry.

We then argue that the true solution of the orbifold
theory has vacua in which the tachyon operator
condenses. We discuss the true vacuum structure of the orbifold theory
and comment on its relevance
for the issue of the closed-string--tachyon condensation in string theory.

The paper is organized as follows. In Sect.~\ref{secone} we show that NPO
requires unbroken \ztwo symmetry and  provide  evidence
that this symmetry is broken. Section~\ref{odpa} is devoted to a
discussion of the
order parameter(s) in the orbifold theory. In Sect.~\ref{sectwo} we discuss
dynamics of  fractional  domain walls. Section~\ref{secthree} is devoted to
low-energy theorems.
Section~\ref{sectype0} discusses the relation between type-0
string theory and the \ztwo orbifold field theory. In Sect.~\ref{orbivsor}
we comment on the difference between orbifold and orientifold daughters.
Finally, in Sect.~\ref{con} we summarize our results and outline
possible issues for future investigations.

After the first version of this paper appeared in the eprint form, a related
work  was submitted \cite{KUY}. We agree with a part
of  criticism presented \cite{KUY}. In particular,
in Ref.~\cite{GS} and in the first version of this paper
low-energy theorems were used to discriminate
between the parent and orbifold daughter theories.
These theorems become instrumental
under the assumption of  coincidence
between the corresponding vacuum condensates.
The vacuum condensate coincidence, imposed previously,
is seemingly not necessary and,
in fact, does not hold in a toy model
we have recently
analyzed. Relaxing this requirement makes
the above low-energy theorems (and gravitational anomalies)
uninformative. If one allows for unequal condensates, they
cannot be used to prove (or disprove) that the \ztwo symmetry is broken.
We revised the manuscript accordingly.

We
strongly disagree, however, with the analysis
of the domain wall issue presented in \cite{KUY}.
Domain walls dynamics in \ztwo orbfiold field theories is
incompatible with planar equivalence.

\section{The role of \ztwo in the proof of planar equivalence}
\label{secone}

Let us analyze whether or not nonperturbative
planar equivalence takes place in the orbifold
theory, following the line of reasoning established in
\cite{ASV-two}. We refer the reader to this paper for a detailed discussion of the procedure.

\begin{figure}[ht]
\centerline{\includegraphics[width=2.5in]{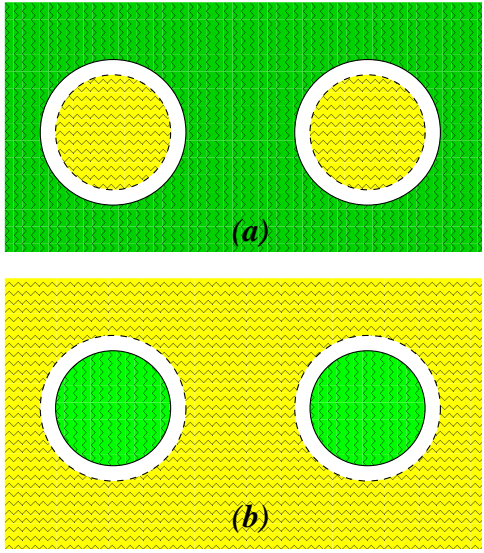}}
\caption{\footnotesize The fermion loop expansion of the partition
function at $N\to\infty$. }
\label{ORBone}
\end{figure}

Here we will consider,   as a particular example,
a two-fermion-loop contribution (see Fig.~\ref{ORBone})
to the partition function.

Each fermion loop consists of two lines: one solid and one dashed.
The solid line denotes propagation of the (fundamental) color index belonging to the ``electric" SU$(N)$ while the dashed  line denotes propagation of the  color index belonging to the ``magnetic" SU$(N)$. ``Electric" and ``magnetic" gluon fields are marked in Fig.~\ref{ORBone}
by  vertical and horizontal shadings, respectively.

Each fermion loop represents, in fact,
\beq
\label{defGa}
\Gamma  [A]=\log \, \det \left( i\not\! \partial + \not\!\!  A ^a
\, T^a   - m \right) \, .
\eeq
A mass term is introduced for regularization. It is very important
(see below). In what follows we will assume for definiteness
that $m$ is real and $m>0$. We note that
\beq
A\, T =A_e\, T_e + A_m\, T_m\,.
\label{cg}
\eeq

\newpage

Moreover,
\beqn
\label{wlineint}
\Gamma _r [A, J_\Psi] &=&
-{1\over 2} \int _0 ^\infty {dT \over T} \nonumber\\[3mm]
&\times&
\int {\cal D} x {\cal D}\psi
\, \exp \left\{ -\int _{\epsilon} ^T d\tau \, \left ( {1\over 2} \dot x ^\mu \dot x ^\mu + {1\over
2} \psi ^\mu \dot \psi ^\mu -{1\over 2} m^2  \right )\right\} \nonumber \\[3mm]
&\times &  {\rm Tr }\,
{\cal P}\exp \left\{   i\int _0 ^T d\tau
\,  \left (A_\mu ^a \dot x^\mu -\frac{1}{2} \psi ^\mu F_{\mu \nu} ^a \psi ^\nu
\right ) T^a  \right\}  \, ,
\eeqn
with the same convention regarding  $AT$ and $FT$ as in Eq.~(\ref{cg}).
In the above expressions the gluon field is considered as
background.
Averaging over the vacuum gluon field is
performed at the very end.

The requirement that the fermion loops are connected
through the gluon field enforces that only selected
contractions are possible in the orbifold theory.
In particular on the diagrams of Fig.~\ref{ORBone}
the outside loops must be {\em both} either solid lines
or dashed lines ($a$, $b$, respectively). Solid-dashed combination is excluded as it represents a disconnected graph. In the parent SYM theory we deal with a single SU($2N)$,
and all contractions are possible.

In perturbation theory the contributions
from the diagrams \ref{ORBone}$a$ and \ref{ORBone}$b$
are equal. The combinatorics is such that adding up  \ref{ORBone}$a$ and \ref{ORBone}$b$ one exactly reproduces
the two-fermion-loop contribution in SYM theory
provided one performs the following coupling rescaling:
\beq
g^2_D = 2 \, g^2_P\,,
\label{ccr}
\eeq
where $P$ and $D$ stand for the parent and  daughter (orbifold) theories.
The above rescaling ensures that the 't Hooft couplings
are the same in the parent and daughter theories.

In the perturbative planar equivalence --- a solidly established fact ---
the vacuum angle $\theta $ plays no role since it does not show up
in perturbation theory. A correspondence between parent and daughter $\theta's$
following from NPO can be derived from holomorphic dependences of the bifermion condensates
on the complexified coupling constants. If
the vacuum angles in the parent and orbifold daughter theories are introduced as
\beqn
\Delta {\cal L}_P &=& \frac{\theta_P}{32\pi^2}\,  F_{\mu\nu}^a \tilde F^{\mu\nu\,,a }\,, \nonumber\\[3mm]
\Delta {\cal L}_D &=&\frac{\theta_D}{32\pi^2}\,\, \sum_{\ell =e,m} F^a_{(\ell)\mu\nu} \tilde F_{(\ell)}^{\mu\nu ,\,a}\,,
\label{tapd}
\eeqn
then it is not difficult to show
that the vacuum angles
must be rescaled too \cite{en,GS,Yaffe},
\beq
\theta_D = \frac{1}{2} \, \theta_P\,.
\label{tar}
\eeq

Equations (\ref{ccr}) and (\ref{tar})
are equivalent to the statement of correspondence between the
holomorphic coupling constants.

\section{Planar equivalence: what does it mean?}
\label{pewdim}

In establishing large-$N$ equivalence between distinct theories, with distinct vacuum
structure (and, as we will see shortly, the vacuum structure of the parent theory is
not maintained upon projection to the orbifold daughter theory)
we must carefully specify what this equivalence might actually   mean.
Any theory is characterized by a set of physical quantities that scale differently in the 't Hooft limit.
For instance, particle masses are assumed to be $N$-independent,
their residues in appropriate currents grow with $N$, particle widths fall with $N$
(so that at $N\to\infty$ all mesons are stable), the vacuum energy density scales as $N^2$,
the number of vacua scales as $N^1$,
and so on. When two theories with with distinct vacuum
structure are compared (but with the
same  scale
parameter $\Lambda$), physical equivalence of the theories in question
need not necessarily mean full general equality of all $n$-point functions, since such equality
may come into contradiction with appropriate scaling laws.
In particular, some vacuum condensates and  low-energy theorems can be
sensitive to the number of
fundamental degrees of freedom (cf. the $\pi^0\to 2\gamma$ constant
whose consideration led to the conclusion of three colors in {\em bona fide}
QCD in the early 1970s).

It is clear that a minimal requirement
of  planar equivalence is coincidence of the particle spectra in the common sector.
More precisely, let us consider a vacuum $V_P$ of the parent theory that can be mapped
onto a vacuum $V_D$ of the daughter theory and vice versa. We must verify that
both $V_P$ and $V_D$ are stable vacua. If the spectra of particle excitations in both vacua
in the common sector coincide up to $1/N$ corrections we can speak of planar equivalence.

Besides particle excitations the parent and daughter theories may (and do) support
extended excitations, such as domains walls. To consider domain walls
we must consider pairs of vacua $V_P,\,\,V_P^\prime$
and $V_D,\,\,V_D^\prime$ which can be mutually mapped.
Correspondence between the parent and daughter walls
can be included in the requirement
of  planar equivalence.

In the remainder of this section we show that evident distinctions
in the vacuum structure of the parent SYM theory and its orbifold daughter
lead, with necessity,  to a mismatch in certain correlation functions.
In particular, $\theta$ dependences cannot match.
This does not necessarily mean a mismatch in the particle spectra
since the composite
particle masses acquire a dependence on the $\theta$ parameters introduced in Eqs. (\ref{tapd})
(at $m\neq 0$, where $m$ is a small fermion mass term, see below)
only in subleading order in  $1/N$.

The functional-integral representation for the partition function, with the fermion determinant included,
is ill-defined unless we regularize the determinant.
The infrared regularization is ensured by the introduction of a
(small) mass term $m$.
Simultaneously, this  mass term lifts the vacuum degeneracy,
eliminating further ambiguities in  the functional integral.

\begin{figure}[ht]
\centerline{\includegraphics[width=3.5in]{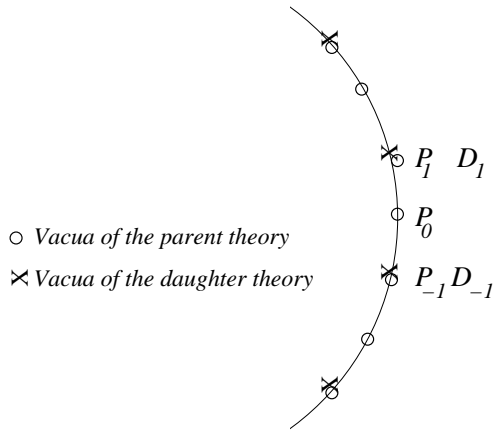}}
\caption{\footnotesize
The vacuum structure in the SU($2N$) SYM theory and its SU$(N)\times$SU($N$) orbifold daughter on the complex plane of the order parameter
(the bifermion condensate $-\langle
\lambda^{a}_{\alpha}\lambda^{a\,,\alpha}
\rangle$ or $-\left\langle
\bar\Psi\left( 1-\gamma_5\right)\Psi \right\rangle$, respectively).}
\label{ORBtwo}
\end{figure}

Let us dwell on the vacuum structure of the parent and daughter theories.
SU($2N$) supersymmetric gluodynamics has $2N$ vacua
labeled by the order parameter, the gluino condensate,\footnote{The gluino condensate in supersymmetric gluodynamics was first conjectured, on the basis of the value of his index, by E.~Witten \cite{wittenindex}.
It was confirmed in an effective Lagrangian approach by G.~Veneziano and S.~Yankielowicz \cite{vy}, and exactly
calculated  by  M.~A.~Shifman and A.~I.~Vainshtein \cite{SV}.
The exact value of the coefficient $-12N$ in Eq.~(\ref{gluco})
(for SU($2N$)) can be extracted from several sources.
All numerical factors are carefully collected for SU(2) in the review paper
\cite{Shifman:1999mv}.  A weak-coupling calculation for SU($N$)
with arbitrary $N$ was carried out in \cite{Davies:1999uw}.
Note, however, that an unconventional definition of the scale parameter $\Lambda$ is used in Ref.~\cite{Davies:1999uw}. One can pass to
the conventional definition of $\Lambda$ either by normalizing the result
to the SU(2) case \cite{Shifman:1999mv} or by analyzing the context of Ref.~\cite{Davies:1999uw}. Both methods give one and the same result.}
\beq
\langle
\lambda^{a}_{\alpha}\lambda^{a\,,\alpha}
\rangle = -6 (2N)\, \Lambda^3 \exp \left(i\, {\frac{2\pi  k +\theta_P}{2N}}\right)\,,
\,\,\, k = 0,1,..., 2N-1\,,
\label{gluco}
\eeq
Its SU$(N)\times$SU($N$) orbifold daughter has
$N$ vacua (under the assumption that \ztwo is unbroken)
which are labeled by the order parameter
\beq
\left\langle
\frac{1}{2}\, \bar\Psi\left( 1-\gamma_5\right)\Psi \right\rangle \sim N\,\Lambda^3 \exp \left(i\, {\frac{2\pi  k +\theta_D}{N}}\right)\,,
\,\,\, k = 0,1,..., N-1\,,
\label{dgluco}
\eeq
Consider an instructive example, namely,
\beq
\theta_D = \pi\,,\qquad \theta_P= 2\pi\,,
\label{ili}
\eeq
and $m$ real and positive,
the vacuum structure is depicted in Fig.~\ref{ORBtwo}. $P_0$
is the unique vacuum of the SYM theory, while $D_{\pm 1}$ are the vacua of the orbifold theory. Note that at $m>0$ the ``vacua" $P_{\pm 1}$ are in fact excited   (or quasistable) because
$$
{\cal E}_{P_{\pm 1}} > {\cal E}_{P_{0}}\,.
$$
The daughter theory has two-fold degeneracy, $$
{\cal E}_{D_{+ 1}} ={\cal E}_{D_{- 1}},
$$
a phenomenon well-known
at $\theta =\pi$. This is the so-called Dashen
phenomenon \cite{dashen},
with all ensuing consequences. Let us emphasize that   physics
at $\theta =\pi$ and $\theta =0$ are
essentially different. In particular, at $\theta =\pi$
spontaneous breaking of discrete symmetries (such as $P$-invariance)
typically occurs \cite{dashen}.

One can consider another instructive example,
$\theta_P =\pi$. At this point, the Dashen phenomenon occurs in the
parent theory. There is a double-fold vacuum degeneracy. At $m\neq 0$
domain walls are unstable, generally speaking. However, a stable domain walls
emerges in the Dashen point, as is well-known from the past.
At the same time, the corresponding vacuum angle of the daughter theory
at this point is $\theta_D =\pi /2$.
The Dashen point is not yet reached. It is clear that there is no equivalence in this aspect.

Coincidence of the vacuum structure in
the parent and daughter theories at  $N =\infty$
implies, generally speaking,  a much broader understanding of planar equivalence.
This is the case for orientifold daughters.
For orbifold daughters one has to stick to the minimal requirement specified in the beginning of this section.

\section{Order parameters}
\label{odpa}

\begin{figure}[ht]
\centerline{\includegraphics[width=3in]{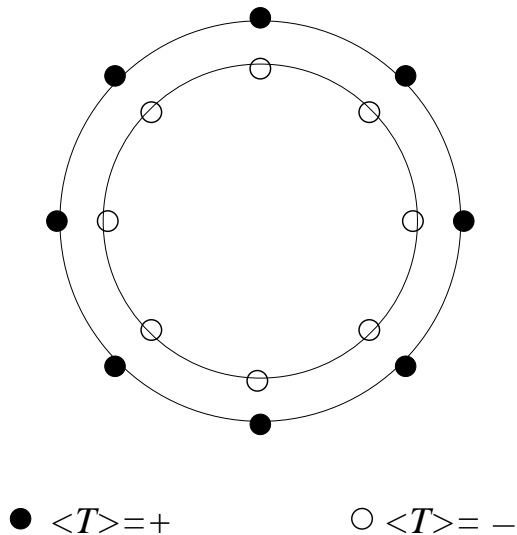}}
\caption{\footnotesize The vacuum structure of the SU(8)$\times$SU(8) orbifold theory. }
\label{biman}
\end{figure}

We will pause here to discuss appropriate order parameters.
In the parent SYM theory the order parameter is the gluino condensate
(\ref{gluco}). In the daughter theory with the spontaneously broken $Z_2$
the bifermion condensate (\ref{dgluco}) is insufficient
for differentiation of all $2N$ vacua of the theory because
it is $Z_2$-even. We must supplement it by a $Z_2$-odd expectation value
of (\ref{iman}). This vacuum expectation value (VEV) is dichotomic.
The bifermion condensate (\ref{dgluco}) in conjunction with
$\langle T\rangle =\pm \Lambda^4$ fully identifies each of the $2N$ degenerate vacua
of the orbifold theory. Somewhat symbolically the vacuum structure
is presented in Fig.~\ref{biman}. The angular
coordinate represents the phase of  (\ref{dgluco}), while the radial coordinate
can take two distinct values
representing the  dichotomic parameter $\langle T\rangle$.

It is instructive to discuss here the $Z_2$-even gluon condensate
$\langle F_e^2 +F_m^2\rangle$. This operator is related to the
total energy-momentum tensor of the theory,
\beq
\theta_\mu^\mu =  - \frac{3N }{32\pi^2}
\sum_{\ell = e,m}\left( F_{\mu\nu}^a\, F_{\mu\nu}^a\right)_\ell\,.
\label{beiman}
\eeq
Since all $2N$ vacua are degenerate, at first sight
the gluon condensate  is no order parameter,
since the VEV of (\ref{beiman}) is the same in all vacua.
Ever since the gluon condensate was introduced in non-Abelian gauge theories \cite{SVZ} people tried
to identify it as an order parameter. In a sense, in the case at hand it is\,!

To be more precise, a nonvanishing (in the {\em planar approximation})
VEV $\langle F_e^2 +F_m^2\rangle \neq 0$
signals the spontaneous breaking
of $Z_2$. Indeed, if $Z_2$ is unbroken
$ \langle F_e^2 +F_m^2\rangle$ reduces to
$ \langle F^2_{\rm SYM}\rangle$.
The latter condensate vanishes due to  supersymmetry of the parent theory.
Hence, the $Z_2$ symmetric vacua in the
daughter theory would have vanishing vacuum energy density.
Since the $Z_2$-symmetric point is unstable,
the {\em bona fide} $Z_2$-asymmetric vacua must have a negative energy density.
Equation (\ref{beiman}) implies then that in the genuine $Z_2$-broken vacua
\beq
\langle F_e^2 +F_m^2\rangle \neq 0\,.
\label{gdesim}
\eeq
Thus, in the case at hand the gluon condensate does play the role of an order parameter, much
in the same way as $ \langle F^2_{\rm SYM}\rangle$
is the order parameter for SUSY breaking in SUSY gluodynamics.
Note that for this reason
$\langle F_e^2 +F_m^2\rangle$ must vanish in (planar)
perturbation theory. {\em Nonperturbatively},\footnote{In the orientifold theory
the gluon condensate vanishes at the leading order. A nonvanishing gluon condensate
at subleading order  was detected in the orientifold theory in Ref.~\cite{ss}.}
Eq.~(\ref{gdesim}) must hold at $O(N^2)$. This prediction from  the broken
$Z_2$ symmetry imposes a strong restriction on the low-energy effective
action for the orbifold daugther. In particular it disfavours
the action suggested in \cite{vafa}.

\begin{figure}[ht]
\centerline{\includegraphics[width=2.5in]{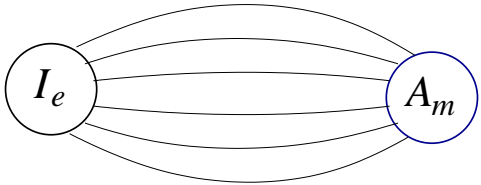}}
\caption{\footnotesize Topologically stable instanton-antiinstanton
pair in the orbifold theory. Instanton belongs to the electric SU($N$)
while antiinstanton to the magnetic SU($N$).} \label{bbiman}
\end{figure}

Needless to say, revealing dynamical distinctions
leading to vanishing/non\-vanishing of $ \langle F^2\rangle$
in the parent/daughter theory is of paramount importance.
We are far from understanding these mechanisms.
We would like to make a single remark regarding instantons,
the only well-studied explicit examples
of nonperturbative field configurations. In the SYM theory instanton does not contribute to the vacuum energy
because of the fermion zero modes (an instanton-antiinstanton configuration
could contribute but it is topologically unstable.)
The orbifold theory exhibits a new phenomenon  (to the best of our knowledge, for the first time ever): topologically {\em stable instanton-antiinstanton
pair}, connected through fermion zero modes, see Fig.~\ref{bbiman}.
The stability is due to the fact that they belong to distinct gauge factors.
Therefore, although the overall topological charge vanishes
(all fermion zero modes are contracted), still
instanton$_e$ cannot annihilate antiinstanton$_m$.

\section{Domain wall dynamics in orbifold field theory}
\label{sectwo}

In this section we discuss the dynamics of domain walls in the \ztwo
orbifold field theory. We discuss both four-dimensional and world-volume
dynamics. Since domain walls are ``QCD D-branes" \cite{witten1997}
the similarity between wall dynamics and D-brane dynamics is clear.
In Sect.~\ref{sectype0} we will discuss the dynamics of D-branes
in type-0 string theory. We will identify the domain walls of the
orbifold daughter theory with the fractional D-branes of type-0 string theory.

Why domain walls ? As well-known, the occurrence of domain walls
is the physical manifestation of spontaneously broken discrete symmetries.
Since our considerations aims at exploring the
\ztwo breaking in the orbifold daughter theory, an analysis of the domain
walls is relevant.

In addition, we will discuss the role of the fractional domain walls of the orbifold theory as fundamental (or constituent) domain walls of the theory in its true vacuum.

\subsection{Four-dimensional perspective}
\label{fdp}

Let us consider the domain walls in the \ztwo orbifold field theory. It is
a SU$_e(N)\times$ SU$_m(N)$ gauge theory with a bifundamental Dirac fermion.
The theory has a global U$_{\rm A}(1)$ axial anomaly  analogous to the U$_{\cal R}(1)$ anomaly in the parent SYM theory.
On the basis of the U$_{\rm A}(1)$ anomaly
one can deduce that the daughter theory
has $N$ degenerate vacua marked by distinct values of
the bifermion condensate $\left\langle
\bar\Psi\left( 1-\gamma_5\right)\Psi \right\rangle$ (see Fig.~\ref{ORBtwo}).
The domain walls can separate these  $N$ vacua. (An alternative terminology:
the domain walls can interpolate between these  $N$ vacua.)
Let us begin with a brief review of the SYM theory domain walls.

The SYM theory contains BPS domain walls \cite{dvali-shifman} that carry both tension $\sigma$ and charge $Q$ (per unit area), with $\sigma = Q$. The expressions for the tension and charge are \cite{armoni-shifman}
\bea
&&
\sigma = {3(2N) \over 32 \pi ^2}\, \int_{\rm wall} \, dz\, {\rm Tr}\, F^2\,, \label{sigmas}\\[2mm]
&&
Q= {3(2N) \over 32 \pi ^2}\,  \int_{\rm wall} \, dz\, {\rm Tr}\, F\tilde F\,,
\label{sigmaq}
\eea
where $z$ is the direction perpendicular
to the wall plane. Equation (\ref{sigmas}) is a consequence of the scale anomaly.

We can consider, as well, the bound state of $k$ elementary walls. These walls
interpolate between the vacua $i$ and $i+k$.  The {\em exact} tension
for the  $k$-wall configuration is \cite{dvali-shifman}
\beq
\sigma (k) = \Lambda ^3 (2N)^2 \sin {\pi k \over 2N}\, .
\eeq
At  $N\to \infty $ it reduces to \beq
\sigma (k) \rightarrow k \sigma (1)
\,. \eeq
In other words, the walls do not interact as  their
total tension is the sum of  tensions of  $k$ {\em free}
1-walls. Although the walls do interact via the exchange of glueballs, there is a perfect cancellation between the contribution of even- and odd-parity glueballs
\cite{armoni-shifman}.  In  Sect.~\ref{wvdp} we will see, from the world-volume
theory standpoint, that the no-force result is due to bose-fermi degeneracy on the wall.

Now we proceed  to the orbifold daughter. Analogously to the parent SYM theory,
the domain walls of the daughter theory carry both tension and charge
which can be evaluated by using the orbifold procedure.

We obtain the following expressions for the
tension and charge of the orbifold theory
domain walls:\,\footnote{In SYM theory such integrals
are well-defined since $\langle F^2\rangle$ vanishes in
SUSY vacua.  In the orbifold theory this is not the
case, see Sect.~\ref{odpa}. Therefore, the integrals in Eq.~(\ref{tqdt})
require a proper regularization.}
\beqn
\sigma _D &=& {3N \over 32 \pi ^2 }\,  \int_{\rm wall} \,  dz \, {\rm Tr} \,  F_e^2 + {3N \over 32\pi ^2 }\,
\int_{\rm wall} \, dz\,  {\rm Tr}\, F_m ^2 \,,
\label{tqdt} \\[2mm]
Q _D  &=& {3N \over 32\pi ^2 }\,  \int_{\rm wall} \, dz \, {\rm Tr}\left(  F\tilde F\right)_e + {3N \over 32\pi ^2 }\,  \int_{\rm wall} \, dz \, {\rm Tr}\left(  F \tilde F\right)_m\,.
\label{tqdtp}
\eeqn

In a bid to reveal inconsistencies
of the NPO conjecture and preparing for such a demonstration in Sect.~\ref{wvdp},
we will look at the domain walls
from a slightly different angle.
It is suggestive to think of the domain walls of the orbifold
theory as of marginally bound states of {\em fractional} ``electric" and ``magnetic" domain walls,  with the following tensions and charges:
\bea
\sigma_e &=&  {3N \over 32 \pi ^2 }\,  \int \,  dz \, {\rm Tr} \,  F_e^2 \,,  \qquad  \sigma _m =  {3N \over 32 \pi ^2 }\,  \int \,  dz \, {\rm Tr} \, F_m ^2 \,,
\nonumber \\[2mm]
Q_e &=&  {3N \over 32 \pi ^2  }\,  \int \,  dz \, {\rm Tr} \, (F\tilde F)_e \,,
\qquad Q _m =  {3N \over 32 \pi ^2  }\,  \int \,  dz \, {\rm Tr} \,  (F \tilde F)_m\,.
\label{tqdtp2}
\eea
Assuming now  that NPO
is valid, {\em and}  the  \ztwo symmetry is unbroken, i.e.
$
\sigma _e = \sigma _m\,,
$
we get
\beq
\sigma_{e,m} = {1\over 2 }\left(\sigma_{e}+\sigma_{m}
\right) ,
\eeq
i.e., a {\em fractional} amount of tension (in full analogy with the fractional D-branes,
see Sect.~\ref{sectype0}). The tensions of the fractional multi-walls are
\beq
\sigma _{e,m} (k) = {1\over 2}\Lambda ^3 N^2 \sin {\pi k \over N}
\to k\, \sigma_{e,m} (1)\quad{\mbox{at}}\quad N\to\infty\,.
\label{sem}
\eeq
When $k=2$, the statement reduces
to that two parallel  electric domain walls do not interact at  $N=\infty$.
Needless to say, the same is valid for the
magnetic walls.
In Sect.~\ref{wvdp} we will demonstrate, using the world-volume description, that two electric domain walls do interact at $N\to\infty$.

\subsection{World-volume (2+1)-dimensional perspective}
\label{wvdp}

Our discussion of the world-volume domain wall dynamics   in the orbifold daughter is closely related to the situation in the parent SUSY theory.
The world-volume theory for $k$-walls in ${\cal N}=1$ gluodynamics was derived in Ref.~\cite{AV}.  It was shown to be a (2+1)-dimensional $U(k)$  theory with  level-$2N$ Chern--Simons term (for the bulk gauge group
SU($2N$)).  The world-volume theory has (2+1)-dimensional \None
supersymmetry. Note that  ${\cal N}=1$ SUSY in three-dimensional  SU$(N)$  theory is dynamically
broken  \cite{witten3d} at  small values of the coefficient  in front of  the Chern-Simons term, 
$k_{cs}\le N/2$. However
this SUSY breakdown  does not happen on the  world-volume of  multiple domain walls
in the parent theory since in this case $k_{cs}=2N$,  and  gauge group is at 
most SU$(N)$.

The action of the theory is
\beq
S=\int d^3 x \left\{ {\rm Tr} \left (
-{1\over 4e^2} F^2 + {2N\over 16\pi} \epsilon ^{ijk} A^i F^{jk}+
{1\over 2} (D_i \Phi)^2 \right ) + {\rm fermions} \right\}\,.
\eeq
All  fields in the action, including the fermion fields, transform in the adjoint
representation of $U(k)$. For definiteness, we will consider the case $k=2$,
which is in a sense minimal, see Sect.~\ref{emfwfc}.

Now, consider the orbifold daughter theory.
The world-volume theory becomes, by virtue of
the orbifold procedure, a U$_e(1)\times$U$_m(1)$ gauge theory with a neutral scalar
field  and {\em bifundamental}
fermions\,\footnote{The same conclusion about the precise form of the world-volume action can be reached by consideration similar to \cite{AV} for type-0 string theory, see Sect.~\ref{sectype0}.}
\beqn
S  &=&
\int d^3 x \left\{ \sum_{\ell=e,m}\left(  -{1\over 4e^2} F_\ell  ^2 + {N\over 16\pi} \epsilon ^{ijk} A_\ell^i \,
F_\ell ^{jk}+  {1\over 2} (\partial_i\Phi_\ell)^2 \right)\right.
\nonumber\\[3mm]
   &+&\left.  \bar \Psi   \left(\Phi_e - \Phi_m\right) \Psi + ...  \right\}.
\label{WVorbifoldp}
\eeqn
As we will see momentarily, the occurrence of the Yukawa coupling
\beq
\bar \Psi   \left(\Phi_e - \Phi_m\right) \Psi \label{yc}
\eeq
in the daughter theory (with no counterpart
in the parent one)  is a  fact of special importance.

We can give the following interpretation to the above expression.
The daughter wall consists of a sum of  electric  and  magnetic  walls that interact with each other
via the bifundamental fermions. In fact, the  electric  branes can be separated from the  magnetic  branes. To see that
this is the case, note that the Yukawa term (\ref{yc}) in the action \eqref{WVorbifoldp} can make
the bifundamental fermion massive.

Indeed, by giving vacuum expectation values \beq
\langle \Phi_e \rangle = v_e\,, \qquad \langle \Phi_m\rangle =v_m\, ,
\eeq
we generate  a mass $\mu$ for the world-volume fermions,
\beq
\mu  = v_e - v_m\,.
\label{mfwvf}
\eeq
When $\mu \to \infty$ the fermions decouple, and we have two
decoupled U(1)  theories. The interpretation is clear: we can give VEV's and {\em separate} the electric domain wall from the magnetic one. The world-volume theory on the separated  electric (or  magnetic) domain walls
is just a bosonic  U(1)  gauge theory with a level-$N$ Chern--Simons term.
It is not supersymmetric. There is no reason for the wall tension non-renormalization
and the no-force statement.

Let us discuss the force between the two walls. It is done by evaluating the
Coleman--Weinberg potential in the presence of a VEV $v$.
A similar calculation was performed in Refs.~\cite{zarembo,zarembo-tseytlin}. The result is
\beq
V/A \sim \int d^3 k\,  \ln \left(k^2 + e^2 v^2 \right) \sim c_0 \Lambda ^3 + c_1 \Lambda e^2 v^2 - c_2 {e^4 v^4 \over m} + ... \,\, , \label{force}
\eeq
where $c_0,c_1$ and $c_2$ are positive coefficients (independent of $N$), $\Lambda$ is a UV cut-off and $m$
is an IR cut-off (the gauge boson Chern--Simons mass).
We can set $c_0$ and $c_1$
to zero by a fine-tuned renormalization.
However, even after renormalization
a repulsive $v^4$ term remains.
This is not surprising. A necessary condition
for the zero force is a degeneracy between bosons
and fermions in the world-volume
theory. This is achieved in the parent theory,
where the world-volume theory is
\None supersymmetric. However, since the
theory on the electric walls of the daughter
theory is purely bosonic we found a repulsion.
This is in contradiction
with the NPO conjecture.

At the end of Sect.~\ref{fdp} we assumed NPO and we reached the conclusion
that there is no force between two parallel electric walls at  $N\to\infty$. However, the microscopic calculation reveals a different answer. Again, the conclusion
is that the \ztwo symmetry must be broken.

\subsection{The fate of electric and magnetic fractional walls as independent constituents in the true solution}
\label{emfwfc}

By studying the fractional domain-wall dynamics we arrived at the conclusion that
the \ztwo symmetry is dynamically broken. Moreover,
the gauge theory has \ztwo-odd vacua.
In other words,
the tachyon field potential has a minimum (the tachyon
field is $T= {\rm Tr} \, F_e ^2 - {\rm Tr}\, F_m ^2$).

\begin{figure}
\epsfxsize=4cm
\centerline{\epsfbox{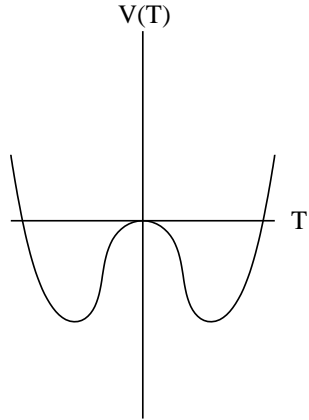}}
\caption{\footnotesize The tachyon field potential. The \ztwo symmetry is dynamically broken in the true vacuum.}
\label{potential}
\end{figure}

The statement that  $V(T)$ is bounded from below is not an assumption --- it can be justified by observing that the regime of large VEV's is fully controlled by semiclassical dynamics. From the field-theoretic standpoint
it is clear that the only possibility open is that in the {\em bona fide} vacuum $\langle T\rangle\sim \Lambda ^4$.
At the same time, non-stabilization of tachyons would mean $\langle T\rangle\gg \Lambda ^4$, which is ruled out.
Therefore,   the tachyon field potential must look like a Higgs
potential, see Fig.~\ref{potential}.

In the parent  \None SYM theory with the gauge group
SU$(2N)$, there are $2N$ vacua, with the gaugino condensate as an order parameter. The $2N$ vacua, being roots of the unity, can be drawn as points on a unit circle,
see Fig.~\ref{ORBtwo}. The domain walls interpolate between the
various vacua.

In the daughter theory the situation is more complicated. Since each vacuum of the $N$ ``false" perturbative vacua splits into two, the vacuum
structure of the gauge theory can be described as {\em two} circles, with
$N$ points on each circle, see Fig.~\ref{biman}.

The wall inheritance from the parent to the daughter theory
proceeds as follows. We first pretend that the daughter theory is planar-equivalent to SYM, and that the \ztwo symmetry is unbroken. Then we must
start from a
2-wall in the parent theory; and it will be inherited, as  the {\em minimal}
wall in the daughter theory.
Indeed, if \ztwo is unbroken there are only $N$ vacua in the orbi-daughter (versus $2N$ in SYM). This is seen from Fig.~\ref{ORBtwo}. If the wall is inherited, the vacua between which it interpolates
must be inherited too. Under NPO only every second vacuum is inherited.
Thus, if we want to consider the wall that is inherited,
we must consider e.g. the wall connecting $D_{-1}$ and $D_{1}$ in the daughter (this is a minimal wall in the daughter),
versus the wall connecting $P_{-1}$ and $P_{1}$ in the parent (this is a 2-wall in SYM).

In the parent theory two 1-walls comprising the 2-wall
do not interact with each other (at $N=\infty$).
If we consider them on top of each other, the world-volume
theory has $U(2)$ gauge symmetry.
However, nobody precludes us from introducing a separation.
Then we will have U(1) on each 1-wall, U(1)$\times$U(1) altogether. The tension of each 1-wall is 1/2
of the tension of the 2-wall, it is well-defined and receives
no quantum corrections.
The fact that the world-volume
theory on each 1-wall is supersymmetric is
in one-to-one correspondence with the absence of quantum corrections.

Now, in the daughter theory, according to NPO, everything should be the same. The minimal wall splits into one
electric and one magnetic
(the electric one connects $D_{-1}$ with the would-be vacuum which is a counterpartner of $P_0$, the magnetic one connects the
would-be vacuum which is a counterpartner of
$P_0$ with $D_{1}$, each having 1/2 of the tension).

However, now the world-volume
theories on e-wall and m-wall are not supersymmetric,
so that there is no reason for the wall tension non-renormaliza\-tion.

In this false orbifold theory, there is also no place
for the ``twisted" walls, since in the false orbifold theory
there are no black vacua and white vacua of Fig.~\ref{biman} ---
supposedly, there is only one per given value
of
$$
\left \langle   \bar\Psi\, \frac{1-\gamma_5}{2}
\Psi
\right \rangle .
$$

A possible visualization of the situation is as follows.
In the parent theory we have degenerate minima at all points
$P_i$. In the true orbifold theory these minima become maxima
(still critical points, but unstable). Near every second maximum two minima develop.
These are true vacua of the true orbifold theory, with $Z_2$ broken.
Of course, the walls that would be inherited from SYM are all unstable,
with tachyonic modes. 1-walls are transformed into electric/magnetic
walls of the orbifold theory, which are still unstable and, in fact, decay.
Each of them separately could decay only into a ``twisted wall"
connecting white and adjacent black true vacua.
The ``untwisted'' electric+magnetic wall can decay into a minimal stable wall of the daughter
theory which connects two neighboring black vacua or
two neighboring white vacua.

\section{D-branes in type-0B string theory}
\label{sectype0}

Orbifold field theories have deep roots in string theory. The particular
case of \ztwo orbifold is related to  type-0 string. In particular,
we can realize the \ztwo orbifold field theory on a brane configuration
which involves D4-branes and orthogonal NS5-branes in type-0 string theory \cite{armoni-kol} (see also \cite{divecchiaetal} for other realizations in type 0B).

Type-0B string theory is a nonsupersymmetric closed string theory, defined by a  diagonal Gliozzi--Scherk--Olive (GSO)
projection that keeps the following sectors:
\beq
(NS-,NS-) \oplus (NS+,NS+) \oplus (R+,R+) \oplus (R-,R-)\,.
\eeq Note the doubling of the R-R fields and the lack of the NS-R sector (closed
string fermions). In addition, it is worth
noting    that the theory contains a tachyon in the
(NS-,NS-) sector.

Due to the doubling of the R-R fields the theory contains {\em two} types
of D-branes, often called  electric  and  magnetic  branes. A combination
of an  electric  and a magnetic brane is referred to as an untwisted
brane. The untwisted brane of the type-0 string is the analogue of the type-II brane. It is useful to think about the electric and magnetic
branes as fractional branes, or as the constituents of the untwisted
brane.

The field theory on a collection of $N$ D-branes of type-II string theory is a supersymmetric U$(N)$ gauge theory.
The field theory on a set of $N$ untwisted D-branes of type-0 string theory
is a U$_e(N)\times$U$_m(N)$ gauge theory with adjoint scalars
and bifundamental fermions \cite{klebanov-tseytlin}. The bosons arise from open strings that
connect electric branes with electric branes or magnetic branes
with magnetic branes. Fermions are due to open strings that connect
electric branes with magnetic branes \cite{bergman}. The situation is depicted in
Fig.~\ref{dbranes} below.

\begin{figure}
\centerline{\includegraphics[width=1.5in]{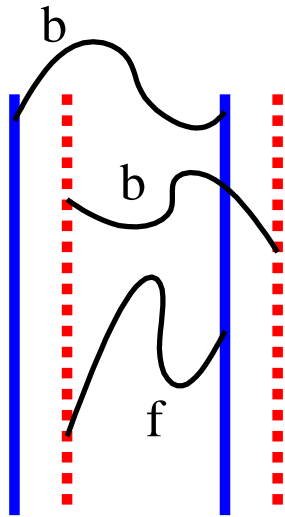}}
\caption{\footnotesize D-branes in type-0 string theory. The open strings that connect
electric branes (solid) with electric branes or magnetic branes (dashed) with magnetic branes are bosons. Open strings that connect electric
branes with magnetic branes are fermions. The field theory on the brane
configuration
is a U$_e(2)\times$U$_m(2)$ gauge theory with bifundamental fermions.
}
\label{dbranes}
\end{figure}

Thus, type-0 string theory provides a  natural framework for discussing
\ztwo orbifold field theories. Indeed, type-0 string theory is a \ztwo orbifold
of type-II string theory.

The forces between D-branes are determined by the annulus diagram. The
short-distance force between a set of
the same-type branes (electric-electric or magnetic-magnetic)
is repulsive \cite{zarembo}. The short-distance force between the opposite
brane pair  (namely, between  electric and magnetic)  is attractive.
The latter matches the picture we presented in Sect.~\ref{emfwfc}.
The forces between untwisted branes, namely between a pair of electric
plus magnetic branes and another such pair is always zero, as in the
supersymmetric theory \cite{zarembo-tseytlin}. The situation is described in Fig.~\ref{annulus} below.

\begin{figure}
\centerline{\includegraphics[width=2.5in]{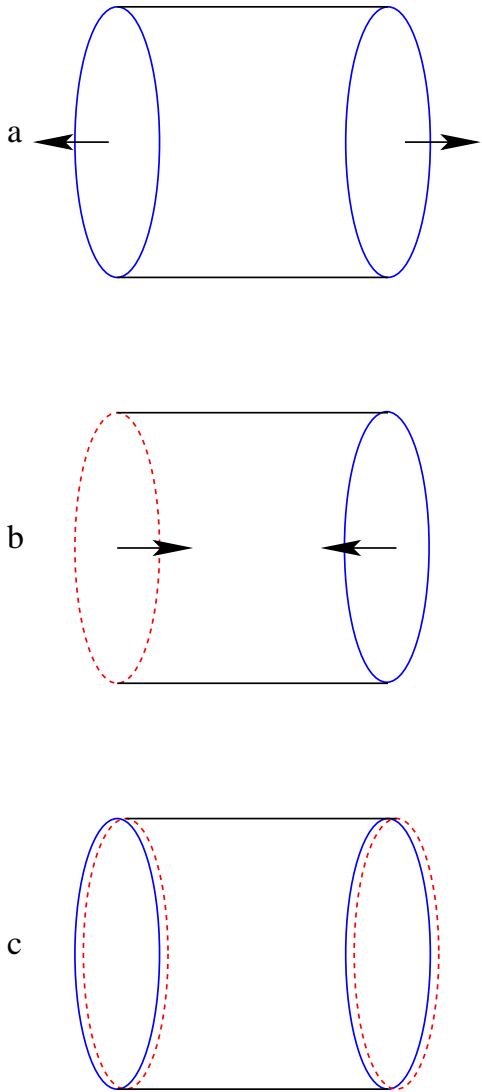}}
\caption{\footnotesize Short distance forces between branes in type-0 string theory: (a) A repulsion between same type branes.
(b) An attraction between opposite type branes. (c) There is no force between untwisted
branes.
}
\label{annulus}
\end{figure}

The above results on the forces between the various D-branes of type-0 string theory can be explained via either the closed string channel or the open
string channel. Let us start with the closed string channel. In order to achieve a zero force between the branes, the attractive
force due to NS-NS modes has to be canceled by the repulsive force
due to R-R modes. Note that the cancellation is among bosons of opposite parity;
it does
not involve fermions (the NS-R sector). No-force
situation is achieved in a SUSY set-up (type II) or for dyonic
(or untwisted) branes of type 0. However, such a cancellation does not occur
in other cases.  The same phenomena can be explained via the open string channel. Here, however, the zero force can be explained due to a cancellation between bosons
(the NS sector) and fermions (R sector). If the world-volume theory on the brane is SUSY (such as type II) or if the spectrum of the modes on the brane is degenerate, the zero force can be achieved.

\subsection{D-branes versus domain walls}
\label{dbvdw}

As has been already mentioned,
Witten  suggested  \cite{witten1997} that domain walls are QCD D-branes. He argued that, since the tension of the domain wall scales as $N\sim g_{\rm st}^{-1}$ and since the QCD string can end on the wall, it is natural to conjecture such
a relation. Moreover, in \cite{witten1997}  Witten described a domain wall as a wrapped M-theory five-brane.   Acharya and Vafa later suggested  \cite{AV} that domain walls correspond to D4-branes wrapping an $S^2$. By using this realization Acharya and Vafa were able to determine the world-volume theory.

Following \cite{AV} we suggest that the domain walls of the \ztwo orbifold field theory correspond to   various branes of type-0 string theory. This is a very natural proposal, since the four-dimensional orbifold field theory itself
can be realized
on a collection of D-branes of the type-0 string \cite{armoni-kol}.

We suggest the following: an electric domain wall corresponds to an electric
brane, a magnetic domain wall corresponds to a magnetic brane and, finally,
the untwisted domain wall --- a pair of electric and magnetic
domain  walls --- corresponds to an untwisted brane.

By using the above identifications and \cite{AV} we can get the world-volume
theory on  various domain walls. The answer follows from an analysis of the
annulus diagram in type-0 string theory \cite{bergman}. For $k$ coincident electric (or magnetic) domain walls it is a (2+1)-dimensional U$(k)$ gauge
theory with a real scalar in the
adjoint representation and a level-$N$ Chern--Simons term. The theory on the collection of $k$ untwisted domain walls
is a U$_e(k)\times$U$_m(k)$ gauge theory
with a real scalar in the adjoint representation of each gauge factor,
a Chern--Simons term
for each factor and bifundamental fermions.

\subsection{Closed-string tachyon condensation}
\label{cstc}

In the previous sections  the breaking of the \ztwo symmetry in the orbifold field theory was proven, mostly within field-theoretical framework,
and consequences outlined. An obvious order parameter
for the \ztwo-symmetry breaking is the tachyon operator
$T \equiv {\rm Tr}\, F_e ^2 - {\rm Tr}\, F_m ^2,\,$
see Eqs. (\ref{iman}), (\ref{cotf}).
The field-theory analysis suggests that $T$  acquires a VEV dynamically and develops a potential of the Higgs type (see Fig.~\ref{potential}).

This conclusion is actually very natural, once the relation with type-0
string theory is established.
If the orbifold field theory is dual to type-0B string theory on a certain manifold (Maldacena--Nu\~nez \cite{MN}, Klebanov--Strassler \cite{KS}, or
$\, C^3/Z_2\times Z_2\,\,$ \cite{divecchiaetal}) then by the
operator/closed-string relation of the AdS/CFT
correspondence, the operator ${\rm Tr}\, F_e ^2 - {\rm Tr}\, F_m ^2$ couples to the tachyon mode of the type-0 string \cite{klebanov}.

It is also clear that,
if there is a duality between a tachyonic string theory and a gauge theory,
then the gauge theory must suffer from an instability at strong coupling \cite{klebanov}.\footnote{The problems due to the tachyonic mode are
solved in orientifold field theories for the simple reason that the dual string theory is nontachyonic, see Sect.~\ref{orbivsor}.}

The situation, however, is not so simple. The tachyon mode has a negative mass square
\beq
m^2 = -{2\over \alpha'}
\eeq
on a flat background, at tree level. The curvature or R-R flux
or, maybe, sigma-model corrections can create a potential for the tachyon. It is very difficult to answer the question of the fate of the closed-string tachyon, especially when the theory is compactified on a non-flat manifold and in the presence of
the R-R flux.

In the case of the \ztwo daughter of ${\cal N}=4$ SYM theory,
which was conjectured
to be dual to type-0B string theory on $AdS _5 \times S^5$ background, Klebanov and Tseytlin \cite{klebanov}
argued that the tachyon mass will be shifted toward positive values, when the 't Hooft coupling is smaller than a certain critical value, namely,
\beq
\alpha ' m^2 = -2 +{c \over \sqrt \lambda} \, .
\label{tachyonmass}
\eeq
However, the full potential for the tachyon field at {\em strong} 't Hooft coupling remained unknown.
Our field-theory analysis suggests a definite
answer to this question. We argue that, if there is a type-0 string model
which is dual to the \ztwo orbifold theory,
the potential for the tachyon mode is as shown in Fig.~\ref{potential}.

We conclude this section by quoting A. M. Polyakov  \cite{polyakov} who discussed the fate
of the tachyon of noncritical type-0 string theory is his paper {\em The Wall of The Cave}:

\begin{quotation}

\noindent
{\sl Presumably, this tachyon should be of the ``good'' variety and peacefully condense in the bulk. }

\end{quotation}

\section{Orbifolds versus orientifolds}
\label{orbivsor}

In this short section, we would like to explain the conceptual difference between
orbifold field theories  and  orientifold field theories, or why
the conjectured planar equivalence \cite{strassler}
does hold for the orientifold field theories \cite{ASV-two}
and fails for orbifold ones.

Let us start with orbifold theories.
The bold conjecture relates supersymmetric theories with the untwisted
sector of the orbifold daughter. It does not address
the twisted sector of the gauge theory but {\it assumes}
that the twisted sector is ``kosher" (or that it decouples from  dynamics of the untwisted sector).

However, as was already argued in the present and previous works,
a necessary condition for
nonperturbative planar equivalence between a supersymmetric
theory and a nonsupersymmetric orbifold daughter, is that the
daughter theory inherits the SUSY vacua.

In this paper we demonstrated that a condensate (\ref{cotf}) develops, hence the vacuum structure
of the orbifold theory is different
from that of the parent SUSY theory.
Multiple evidence for the condensate \eqref{cotf} was obtained,
in particular,  via investigation of fractional domain walls.

String theorists are familiar with this phenomenon. Type-II strings on orbifold singularities of the form $\,C^3/ Z_n\, $, or type-0 strings {\it always}
contain a tachyon in the twisted sector (and fractional branes).

For orientifold theories the situation is conceptually different. This
nonsupersymmetric gauge
theory does not contain a twisted sector and, in particular, it does not contain fractional domain walls; hence, it is guaranteed that the
theory inherits its vacua from the SUSY parent.

Similarly, the candidate for a string dual of the orientifold theory ---
Sagnotti's type-0' model \cite{sagnotti} --- does not contain a tachyon since it was projected out by  orientifolding.

Thus, either from the string-theory side or from the field-theory side,
it is evident that a tachyon-free model is a much better starting point
for the investigation of nonsupersymmetric gauge (or string) dynamics.

\section{Conclusions}
\label{con}

The goal of this work is to determine the vacuum
structure of a non-super\-symmetric gauge theory,
the orbifold theory. The problem is extremely difficult,
since the answer lies in the nonperturbative
regime of the theory.
The \ztwo orbifold theory is obtained
from \None SUSY gluodynamics by
orbifolding. Nonperturbative planar equivalence
for such daughter-parent pairs
(SUSY--non-SUSY) was suggested in Ref.~\cite{strassler}.
While
nonperturbative planar equivalence
was proven for orientifold daughter \cite{AAMSGV,ASV-one,ASV-two},
with multiple consequences that ensued almost immediately,
theorists continued working on orbifold daughters.
Evidence reported in this paper points to
nonperturbative nonequivalence. Of course,
one can say that on the positive side nonperturbative nonequivalence
implies spontaneous breaking of
$Z_2$ of the orbifol daughter.

Our investigation suggests a different picture in the orbifold case.
Based on domain wall dynamics  we  arrived at Fig.~\ref{biman}.
$N$ ``pre-vacua" that could be inferred from the chiral condensate
(\ref{dgluco}), due to \ztwo breaking,   split  into $2N$ vacua,
$N$ ``white" and $N$ ``black."
Each vacuum is uniquely parametrized by two order parameters: the bifermion condensate and the tachyon vacuum expectation value
(\ref{cotf}). In a theory with multiple vacua,  interpolating domain walls
of distinct types exist. In the true orbifold solution we have walls
connecting two white vacua (or two black ones)
which can be interpreted as a bound state of an electric plus magnetic wall pair,
each of these e,m-walls being individually unstable. We also have twisted walls
interpolating between a black vacuum and a white one.

Several possible directions of future research are
at the surface.  It would be interesting to investigate other
gauge theories applying the set of tools used in the
present work.
An  interesting question is whether there exists at all
a daughter non-SUSY orbifold theory whose
the vacuum structure {\it is} inherited from the parent SUSY theory.\footnote{A good starting point could
be a nonsupersymmetric, tachyon-free, closed-string theory. To the best of our  knowledge, the only two examples of non-SUSY tachyon-free string theories
are type-0' and the $SO(16)\times SO(16)$ heterotic strings. Presumably, they are  dual to each other.} Another possible line of investigation is a
derivation of an effective Lagrangian of the
Veneziano--Yankielowicz type \cite{vy}
that would generalize the Lagrangians
\cite{ss} and \cite{MS} to include  effects due to  spontaneously
broken $Z_2$.

Examples of cross-fertilization between string theories and gauge field theories are abundant. In the recent years the direction ``from fields to strings"
is becoming increasingly useful. The present work suggests
another topic along these lines: studying
closed-string tachyon condensation basing on the analogous phenomenon
in field theory. Detailed analysis of the supergravity solutions
corresponding to the strong-coupling limit of the orbifold
daughter theory could shed light on these issues.

\section*{Acknowledgments}
\addcontentsline{toc}{section}{Acknowledgments}

We thank Arkady Tseytlin for valuable comments on the manuscript.
The work  of M.S.  was supported in part by DOE grant DE-FG02-94ER408. The work of A.A. was supported by a PPARC advanced fellowship. The work of A.G. was supported in part by the French-Russian Exchange Program,
CRDF grant RUP2-261-MO-04,
and RFBR grant 04-011-00646.  A.G. is grateful to  LPT at
Universit\'e Paris XI,
where a part of the work was done, for kind hospitality, and acknowledges
a discussion with L. Yaffe. We also thank our PRD referee for useful
comments.

\section*{Appendices}
\addcontentsline{toc}{section}{Appendices: Low-energy theorems}

\renewcommand{\thesection}{A}
\setcounter{section}{0}

\renewcommand{\theequation}{A.\arabic{equation}}
\setcounter{equation}{0}

\section{Trace anomaly low-energy theorems}
\label{secthree}

Here we will derive and discuss low-energy theorems related to the trace anomaly.

Let the parent theory be ${\cal N}=1$ SUSY Yang--Mills
theory with SU($2 N$) gauge group,
\beq
{\cal{L}}=  \frac{1}{g^2_P}\left(- \frac{1}{4}\,  F_{\mu\nu}^a F^{\mu\nu\,,a }+
i \bar{\lambda}_{\dot\alpha}^a {\cal D}^{\dot\alpha\alpha} \lambda_{\alpha}^a
\right)\, +\frac{\theta_P}{32\pi^2}\,  F_{\mu\nu}^a \tilde F^{\mu\nu\,,a } ,
\label{one}
\eeq
where $\lambda_{\alpha} $ is the Weyl spinor in the adjoint.
The theory has ${2 N}$
chiral-asymmetric vacua labeled by the value of the gluino condensate
$\langle \lambda\lambda\rangle$. We will have to add a small gluino mass term,
which will lift the vacuum state from zero and break SUSY, and will
make $\theta$ dependence physical and observable.

The daughter theory
is the gauge theory with $\mbox{SU}(N)\times \mbox{SU}(N) $ gauge
group,  two Weyl bifundamentals and the rescaling law
(\ref{ccr}), (\ref{tar}).
The Lagrangian of the daughter theory is
\beq
{\cal{L}}=  \frac{1}{g^2_D}
\left(- \frac{1}{4}\, \sum_{\ell =e,m} F_{(\ell)\mu\nu} F_{(\ell)}^{\mu\nu } +i
\sum_{\ell =e,m}  \bar \chi_{\ell}   {\cal D} \, \chi_{\ell }\right)
+ {\cal{L}}_\theta \, ,
\label{sixp}
\eeq
where the covariant derivative is defined as $D_{\mu}= \partial_{\mu} - i
\sum_{\ell} A_{(\ell) \mu}\,  T^{(\ell)}$, while   $T^{(\ell)}$ are the generators of
the gauge symmetry with respect to the $\ell$-th group SU($N$)  (here $\ell =e,m$).
The fermion fields have the following color assignment
\beq
\chi_1 \to \chi^a_i\,,\qquad \chi_2 \to \eta_b^j\,,
\eeq
where $a,b$ are fundamental/antifundamental
indices belonging to the first  (electric) SU($N$) while
$i,j$ are fundamental/antifundamental
indices belonging to the second  (magnetic) SU($N$).
Then $\chi_1\chi_2\equiv \chi^a_i \eta_a^i  $
is a gauge invariant chiral order parameter. In the theory (\ref{sixp}), using the existence of the above parameter,
we will introduce the fermion mass term
{\em exactly equal to the projection}
of the gluino mass term of the parent theory
into the daughter one.
The daughter theory is non-supersymmetric.

Now, both the parent theory and its orbifold daughter
are endowed with appropriate (small) mass terms
for the fermions. The mass terms are needed for (i) IR regularization;
(ii) making the vacuum energy density  ${\cal E}_{\rm vac} \sim O(N^2)$.
In the massless limit the ${\cal E}_{\rm vac} \sim O(N)$,
and it is very hard to track subleading terms.
(See, however, Ref.~\cite{ss}.) We will discuss
only the terms $\sim O(N^2)$ in ${\cal E}_{\rm vac}$.

Let us use the fact that
\beqn
\theta_\mu^\mu &=& \frac{\beta_0 }{32\pi^2}\,F_{\mu\nu}^a\, F_{\mu\nu}^a + \mbox{ ferm mass term} = 4{\cal E}_{\rm vac}\,;
\\[3mm]
\beta_0 &=& - 6N\,\,\, \mbox{(parent)}; \quad \beta_0= - 3N\,\,\, \mbox{(daughter)}\,.
\eeqn
Combining it with the fact that
${\cal E}_{\rm vac}$ = VEV of the very same mass term, we conclude
that
\beq
- \frac{6N}{2^7 \pi^2}\left\langle F_{\mu\nu}^a\, F_{\mu\nu}^a
\right\rangle  = \frac{3}{4} {\cal E}_{\rm vac} \eeq
in the parent theory, and
\beq
-\frac{3N}{2^7 \pi^2}\left\langle\left( F_{\mu\nu}^a\, F_{\mu\nu}^a\right)_e + \left( F_{\mu\nu}^a\, F_{\mu\nu}^a\right)_m
\right\rangle  = \frac{3}{4} {\cal E}_{\rm vac} \eeq
in the daughter one, and the
vacuum energy density of the parent is twice higher than that of the daughter.

\vspace{2mm}

Let us    make a comment concerning the order parameters
in the daughter theory. From the low-energy theorems \cite{NSVZ}
we get \beqn
&&4\left\langle \left( F_{\mu\nu}^a\, F_{\mu\nu}^a\right)_e - \left( F_{\mu\nu}^a\, F_{\mu\nu}^a\right)_m \right\rangle
=
-\frac{3N}{32\pi^2}\int d^4x
\nonumber\\[3mm]
&&\times \left\{
\left\langle ( F_{\mu\nu}^a\, F_{\mu\nu}^a) (x),\,
(F_{\mu\nu}^a\, F_{\mu\nu}^a) (0)\right\rangle_e -\left\langle  (F_{\mu\nu}^a\, F_{\mu\nu}^a)(x),\, ( F_{\mu\nu}^a\, F_{\mu\nu}^a)(0)\right\rangle_m \right\},\nonumber
\\
\eeqn
which means that the mixed e-m correlator does not contribute
to the condensate of the twisted field. In other words, the
condensate of the twisted field corresponds to a
difference in the  interactions between the ``electric'' and
``magnetic'' domain walls. On the other hand, a similar
low-energy theorem for the ``untwisted'' condensate

\beqn
&&4\left\langle \left( F_{\mu\nu}^a\, F_{\mu\nu}^a\right)_e + \left( F_{\mu\nu}^a\, F_{\mu\nu}^a\right)_m \right\rangle
=
-\frac{3N}{32\pi^2}\int d^4x
\nonumber\\[4mm]
&&\times \left\{
\left\langle ( F_{\mu\nu}^a\, F_{\mu\nu}^a)_e (x)
+(F_{\mu\nu}^a\ F_{\mu\nu}^a)_m (x) , \,\,\, (F_{\mu\nu}^a\, F_{\mu\nu}^a)_e(0)+ ( F_{\mu\nu}^a\, F_{\mu\nu}^a)_m(0)\right\rangle \right\},\nonumber
\\
\eeqn
shows that the mixed e-m correlator contributes in this
case. Let us remark that the mixed instanton-antiinstanton
pairs mentioned in Sect.~\ref{odpa} can contribute to the ``untwisted''
condensate or the vacuum energy only.

\renewcommand{\thesection}{B}
\setcounter{section}{0}

\renewcommand{\theequation}{B.\arabic{equation}}
\setcounter{equation}{0}

\section{Topological susceptibilities}

We define $\theta$ terms in Eqs.~(\ref{tapd}). A few comments on this definition
will be presented shortly. It is important   that (in the parent theory)
\beq
{\cal E}_P = - {\cal E}_{0,P} \cos\frac{\theta_P}{2N}\,.
\label{two}
\eeq
Here ${\cal E}_P $ is the vacuum energy density in the parent theory,
${\cal E}_{0,P}$ is a positive constant (proportional to $m_{\rm gluino}\Lambda^3$).
The $N$ dependence in Eq.~(\ref{two})  follows from Witten-type arguments combined with the fact that there are $2N$ vacua all entangled
in the process of the $\theta$ evolution in the parent theory. This entanglement leads
to apparent periodicity  $2\pi\cdot  2N$ rather than $2\pi$.

Differentiating Eq. (\ref{two}) twice with respect to $\theta_P$,  using Eq. (\ref{one}) and setting $\theta_P=0$
after differentiation we get the following result for the
topological susceptibility in the parent theory:
\beq
{\cal T}_P \equiv   i\int d^4 x
\left\langle \frac{1}{32 \pi^2}\,
F_{\mu\nu}^{a}\tilde F_{\mu\nu}^a (x) ,\,\, \frac{1}{32 \pi^2}\,
F_{\mu\nu}^{a}\tilde F_{\mu\nu}^a (0) \right\rangle_{\theta=0}
=  {\cal E}_{0,P}\,\frac{1}{(2N)^2} \,.
\label{three}
\eeq

Now, let us turn to the daughter theory and discuss the $\theta$ term in the daughter theory. We introduce
$\theta_D$ in such a way that the physical $2\pi$ periodicity
in $\theta_D$  is maintained, as indicated in Eq.~(\ref{tapd}). For convenience we reproduce the appropriate part here,
\beq
{\cal{L}}_\theta =\frac{\theta_D}{32\pi^2}\,\, \sum_{\ell =e,m} F_{(\ell)\mu\nu} \tilde F_{(\ell)}^{\mu\nu }\,.
\label{seven}
\eeq
Equation (\ref{seven}) is  consistent with the physical $2\pi$ periodicity. Indeed, in the daughter theory one can have instanton just in one of the
two SU($N$)'s, with a trivial background in the other
SU($N$). Then, the normalization in Eq. (\ref{seven}) is
standard. It is   consistent with Eqs. (\ref{ccr}) and (\ref{tar}).
This is best seen upon transition to the canonically normalized
kinetic terms in both theories. Indeed, then
$g_P^2\,\theta_P \leftrightarrow g_D^2\,\theta_D$, and consequently Eq. (\ref{ccr}) implies  (\ref{tar}).

Next, let us consider the following two-point function
\beq
\Pi_P (q)=
i\int d^4 x\, e^{iqx}
\left\langle \frac{1}{32 \pi^2}\,
F_{\mu\nu}^{a}\tilde F_{\mu\nu}^a (x) ,\,\, \frac{1}{32 \pi^2}\,
F_{\mu\nu}^{a}\tilde F_{\mu\nu}^a (0) \right\rangle_{\theta=0}
\label{eight}
\eeq
at large $q$ in the parent theory. At $q^2=0$ it reduces to ${\cal T}_P$.

Let us normalize its counterpart in the daughter theory in such a way that at large $q$ (in the perturbative domain)
the corresponding planar graphs are {\em equal}.
It  is not difficult to see that the equal-normalization condition
implies
\beq
\Pi_D (q)
= i\, \frac{1}{2}\, \int d^4 x \, e^{iqx}
\left\langle \frac{1}{32 \pi^2}\, \sum_{\ell =e,m} \,\,F_{\mu\nu}^{a_\ell }
\tilde F^{a_\ell}_{\mu\nu} (x) ,\,\, \frac{1}{32 \pi^2}\,
\sum_{\ell =e,m}\,\,F_{\mu\nu}^{a_\ell }
\tilde F^{a_\ell}_{\mu\nu} (0) \right\rangle_{\theta=0}\,.
\label{nine}
\eeq
Indeed, at large $q^2$ we have
\beq
{\rm Im}\, \Pi_P (q)=  g_P^4 (2N)^2 q^4\,,\qquad {\rm Im}\, \Pi_D (q)= \frac{1}{2} \, 2\,  g_D^2 (N)^2 q^4\,,
\label{ten}
\eeq
where the factor $1/2$ in the last term reflects $1/2$ in the defining equation (\ref{nine}), while $2$ reflects two distinct
SU($N)$ gluons in the daughter theory. Given equation (\ref{ccr}), the perfect match between $\Pi_P (q)$ and $\Pi_D (q)$ at large $q^2$ is achieved, as we intended. From now on we will drop the subscript
$\theta =0$ where it is self-evident.

Now, assume the orbifold planar equivalence holds
nonperturbatively \cite{strassler}. Then we must conclude that
\beq
\Pi_D (q=0) = \Pi_P (q=0)\,,
\label{eleven}
\eeq
implying, in turn, that
\beqn
\frac{1}{2}\, {\cal T}_D &\equiv&
i\, \frac{1}{2}\, \int d^4 x \, \left\langle \frac{1}{32 \pi^2}\, \sum_{\ell =e,m} \,\,F_{\mu\nu}^{a_\ell }
\tilde F^{a_\ell}_{\mu\nu} (x) ,\,\, \frac{1}{32 \pi^2}\,
\sum_{\ell =e,m} \,\,F_{\mu\nu}^{a_\ell }
\tilde F^{a_\ell}_{\mu\nu} (0) \right\rangle
\nonumber\\[3mm]
& = &{\cal T}_P =\frac{{\cal E}_{0,P}}{4N^2}\,.
\label{twelve}
\eeqn

On the other hand, in the daughter theory --- remember it has allegedly $N$ rather than $2N$ vacua --- the dependence of the vacuum energy density on the
$\theta$ angle is as follows:
\beq
{\cal E}_D = - {\cal E}_{0,D} \cos\frac{\theta_D}{N}\,.
\label{thirteen}
\eeq
Differentiating twice over $\theta_D$ and setting $\theta_D=0$,
we obtain
\beq
{\cal T}_D =
\frac{{\cal E}_{0,D}}{N^2}\,. \label{fourteen}
\eeq
Combining now Eqs.  (\ref{twelve}) and  (\ref{fourteen})
we come to the very same conclusion as in Appendix A,
\beq
{\cal E}_{0,P} = 2 \, {\cal E}_{0,D}\,.
\label{fifteen}
\eeq

\renewcommand{\thesection}{C}
\setcounter{section}{0}

\renewcommand{\theequation}{C.\arabic{equation}}
\setcounter{equation}{0}
\section{Gravitational chiral anomalies}
\label{gcam}

The parent SUSY gluodynamics and the daughter orbifold theories
have classically conserved axial currents which are anomalous at the
quantum level. In addition to the gluon anomaly of the Adler-Bell-Jackiw type,
one can consider the gravitational anomaly whose existence was
first noted in \cite{gra1,gra2}.

At first we will have to establish appropriately normalized operators
which are related by the orbifold projection.
If the axial current in the parent theory (in the Weyl
representation) is\,\footnote{The trace is normalized in such a way
that ${\rm Tr}  \, \bar\lambda^{\dot\alpha}\left(\sigma^\mu\right)_{\dot\alpha\alpha}\lambda^\alpha =
\bar\lambda^{\dot\alpha\,\,a}\left(\sigma^\mu\right)_{\dot\alpha\alpha}
\lambda^{\alpha\,\,a}$.}
\beq
A_P^\mu = g_P^{-2}\, \, {\rm Tr}  \, \bar\lambda^{\dot\alpha}\left(\sigma^\mu\right)_{\dot\alpha\alpha}\lambda^\alpha\,,
\label{gdur1}
\eeq
its orbifold counterpart is
\beq
A_D^\mu =  g_D^{-2}\, \,  \bar\Psi\gamma^\mu\gamma^5\Psi \,,
\label{gdur2}
\eeq
where $\Psi$ is the bifundamental Dirac spinor.

With these definitions the chiral gluon anomaly takes the form
\beqn
\partial_\mu\, A_P^\mu &=&\frac{2N}{16\pi^2}\,{\rm Tr}\, F_{\mu\nu}\tilde
F^{\mu\nu}\,,
\label{gdur3}
\\[3mm]
\partial_\mu\, A_D^\mu &=&   \frac{N}{16\pi^2}\left\{{\rm Tr}\, \left(F_{\mu\nu}\tilde F^{\mu\nu}\right)_e
+{\rm Tr}\, \left(F_{\mu\nu}\tilde F^{\mu\nu}\right)_m
\right\}.
\label{gdur4}
\eeqn

Now, let us pass to the gravitational anomalies.
For one Dirac fermion it was calculated in
\cite{gra1,gra2}
\beq
\partial_\mu\, A^\mu = - \frac{1}{192\pi^2}R_{\mu\nu\kappa\lambda}\tilde
R^{\mu\nu\kappa\lambda}\,,
\label{kds}
\eeq
where $R_{\mu\nu\kappa\lambda}$ is the Riemann tensor. For simplicity
we  specified Eq.~(\ref{kds})
to the lowest order in $h_{\mu\nu}$. (Otherwise, one must have the covariant
derivative on the left-hand side). To the  lowest order, the right-hand
side is $O(h_{\mu\nu }^2)$ and
$$
\tilde R^{\mu\nu\kappa\lambda} =\frac{1}{2} \varepsilon^{\mu\nu\rho\sigma}
R_{\rho\sigma}^{\,\,\,\,\kappa\lambda}\,.
$$
Equation (\ref{kds}) assumes the axial current normalized to unity, and the unit
coupling $h_{\mu\nu}\, \theta^{\mu\nu}$.

Let us examine the gravitational anomalies in the parent and daughter theories,
expressing the answer in terms of the right-hand side of (\ref{kds}).
In the parent theory the coefficient is $(1/2)\cdot 4N^2= 2N^2 $,
the factor $1/2$ being associated with the Weyl fermions
in Eq.~(\ref{gdur1}). In the daughter theory the coefficient is
$  N^2$.
This factor is the number of the Dirac degrees of freedom.

\renewcommand{\thesection}{D}
\setcounter{section}{0}

\renewcommand{\theequation}{D.\arabic{equation}}
\setcounter{equation}{0}

\section{Additional low-energy theorems
in the orbifold theory}

The orbifold theory admits a class of low-energy theorems
which have no parallel in the parent SYM theory.
They seem  interesting on their own;
some are presented here with a brief comment.

The orbifold theory has two classically conserved currents,
\beq
V^\mu = \bar\Psi\gamma^\mu\Psi \,,\qquad A^\mu =\bar \Psi\gamma^\mu\gamma^5 \Psi\,.
\label{astwo}
\eeq
The axial current  $A^\mu$ is anomalous and can be
projected onto its counterpart in the SYM theory.
At the same time, the vector current  $V^\mu$ is anomaly free.
It has no projection.

We can couple this current  $V^\mu$ to an external gauge boson, a ``photon."
Then the orbifold theory becomes an
SU$(N)\times$SU$(N)\times$U(1) theory.  The U(1) filed strength tensor
will be denoted by ${\cal F}_{\mu\nu}$.
Consideration of the scale and chiral anomalies in this theory
along the lines suggested in \cite{NSVZ}
provides us with low-energy theorems
for the two-photon couplings to the gluon operators,
namely,
\beq \left\langle 0 \left| 3N\,\frac{1}{2^5\pi^2}\sum_\ell\left(F_{\mu\nu}^a
F^{\mu\nu\,\, a}\right)_\ell\right| 2\,\gamma\right\rangle
= \frac{4N^2}{3}\, \frac{1}{2^5\pi^2} \left( {\cal F}_{\mu\nu} {\cal F}^{\mu\nu}\right)_{2\gamma}
\label{asa}
\eeq and
\beq \left\langle 0 \left| N\,\frac{1}{2^4\pi^2}\sum_\ell\left(F_{\mu\nu}^a
\tilde F^{\mu\nu\,\, a}\right)_\ell\right| 2\,\gamma\right\rangle
= -  2N^2 \, \frac{1}{2^4\pi^2} \left( {\cal F}_{\mu\nu} \tilde{{\cal F}}^{\mu\nu}\right)_{2\gamma}\,,
\eeq where the photons are assumed to be on mass shell and
$(k_1+k_2)^2 \to 0$ (here $k_{1,2}$ are photons' momenta).

Note that unlike QCD the orbifold theory at hand has no composite Goldstone mesons. Therefore, a subtle point in the derivation of the scale anomaly
(\ref{asa}) which was revealed in \cite{LS} does not show up here.


\newpage
\addcontentsline{toc}{section}{References}
\small
\begin{thebibliography}{99}
\itemsep -2pt

\bibitem{AAMSGV}
A.~Armoni, M.~Shifman and G.~Veneziano,
Nucl.\ Phys.\ B {\bf 667}, 170 (2003)
[hep-th/0302163].

\bibitem{ASV-one}
A.~Armoni, M.~Shifman and G.~Veneziano,
Phys.\ Rev.\ Lett.\  {\bf 91}, 191601 (2003)
[hep-th/0307097];
Phys.\ Lett.\ B {\bf 579}, 384 (2004)
[hep-th/0309013];
for a review see {\em ``From super-Yang--Mills theory to QCD: Planar equivalence and its implications,''}
in {\sl From Fields to Strings: Circumnavigating Theoretical Physics},
Eds. M. Shifman, J.~Wheater and A. Vainshtein, (World Scientific,
Singapore, 2004), Vol. 1, page 353
[hep-th/0403071].

\bibitem{ASV-two}
A.~Armoni, M.~Shifman and G.~Veneziano,
Phys.\ Rev.\ D {\bf 71}, 045015 (2005)
[hep-th/0412203].

\bibitem{kachru}
S.~Kachru and E.~Silverstein,
Phys.\ Rev.\ Lett.\  {\bf 80}, 4855 (1998)
[hep-th/9802183].

\bibitem{Lawrence:1998ja}
A.~E.~Lawrence, N.~Nekrasov and C.~Vafa,
Nucl.\ Phys.\ B {\bf 533}, 199 (1998)
[hep-th/9803015].

\bibitem{K}
M.~Bershadsky, Z.~Kakushadze and C.~Vafa,
Nucl.\ Phys.\ B {\bf 523}, 59 (1998)
[hep-th/9803076].

\bibitem{KK}
M.~Bershadsky and A.~Johansen,
Nucl.\ Phys.\ B {\bf 536}, 141 (1998)
[hep-th/9803249].

\bibitem{Kakushadze}
Z.~Kakushadze,
Nucl.\ Phys.\ B {\bf 529}, 157 (1998)
[hep-th/9803214]; 
Phys.\ Rev.\ D {\bf 58}, 106003 (1998)
[hep-th/9804184].

\bibitem{schmaltz}
M.~Schmaltz,
Phys.\ Rev.\ D {\bf 59}, 105018 (1999)
[hep-th/9805218].

\bibitem{armoni-kol}
A.~Armoni and B.~Kol,
JHEP {\bf 9907}, 011 (1999)
[hep-th/9906081].

\bibitem{strassler}
M.~J.~Strassler,
{\em On methods for extracting exact non-perturbative results
in non-supersymmetric gauge theories,}  hep-th/0104032.

\bibitem{GS}
A.~Gorsky and M.~Shifman,
Phys.\ Rev.\ D {\bf 67}, 022003 (2003)
[hep-th/0208073].

\bibitem{dt}
D.~Tong,
JHEP {\bf 0303}, 022 (2003)
[hep-th/0212235].

\bibitem{Yaffe}
P.~Kovtun, M.~Unsal and L.~G.~Yaffe,
{\em Necessary and sufficient conditions for non-perturbative equivalences of large $N_c$ orbifold gauge theories,}
hep-th/0411177.

\bibitem{klebanov-tseytlin}
I.~R.~Klebanov and A.~A.~Tseytlin,
Nucl.\ Phys.\ B {\bf 546}, 155 (1999)
[hep-th/9811035].

\bibitem{klebanov}
I.~R.~Klebanov and A.~A.~Tseytlin,
JHEP {\bf 9903}, 015 (1999)
[hep-th/9901101];
I.~R.~Klebanov,
Phys.\ Lett.\ B {\bf 466}, 166 (1999)
[hep-th/9906220].

\bibitem{KUY}
  P.~Kovtun, M.~Unsal and L.~G.~Yaffe,
{\em Can large $N_c$ equivalence between supersymmetric Yang--Mills theory and its
  orbifold projections be valid?},
 hep-th/0505075.

\bibitem{en}
J.~Erlich and A.~Naqvi,
JHEP {\bf 0212}, 047 (2002)
[hep-th/9808026].

\bibitem{wittenindex}
E.~Witten,
Nucl.\ Phys.\ B {\bf 202}, 253 (1982).

\bibitem{vy}
G.~Veneziano and S.~Yankielowicz,
Phys.\ Lett.\ B {\bf 113}, 231 (1982). 

\bibitem{SV}
M.~A.~Shifman and A.~I.~Vainshtein,
Nucl.\ Phys.\ B {\bf 296}, 445 (1988).

\bibitem{Shifman:1999mv}
M.~A.~Shifman and A.~I.~Vainshtein,
{\em Instantons versus supersymmetry: Fifteen years later,}
in M. Shifman, {\sl ITEP Lectures on Particle Physics and Field Theory},
(World Scientific, Singapore, 1999)
Vol. 2, p. 485-647 [hep-th/9902018].

\bibitem{Davies:1999uw}
N.~M.~Davies, T.~J.~Hollowood, V.~V.~Khoze and M.~P.~Mattis,
Nucl.\ Phys.\ B {\bf 559}, 123 (1999)
[hep-th/9905015].

\bibitem{dashen}
R.~F.~Dashen,
Phys.\ Rev.\ D {\bf 3}, 1879 (1971).

\bibitem{SVZ}
M.~A.~Shifman, A.~I.~Vainshtein and V.~I.~Zakharov,
Nucl.\ Phys.\ B {\bf 147}, 385; 448 (1979).

\bibitem{ss}
F.~Sannino and M.~Shifman,
Phys.\ Rev.\ D {\bf 69} (2004) 125004
[hep-th/0309252].

\bibitem{vafa}
R.~Dijkgraaf, A.~Neitzke and C.~Vafa,
{\em Large-$N$ strong coupling dynamics in non-supersymmetric orbifold field
theories,} hep-th/0211194.

\bibitem{witten1997}
E.~Witten,
Nucl.\ Phys.\ B {\bf 507}, 658 (1997)
[hep-th/9706109].

\bibitem{dvali-shifman}
G.~R.~Dvali and M.~A.~Shifman,
Phys.\ Lett.\ B {\bf 396}, 64 (1997)
(E)   B {\bf 407}, 452 (1997)
[hep-th/9612128].

\bibitem{armoni-shifman}
A.~Armoni and M.~Shifman,
Nucl.\ Phys.\ B {\bf 670}, 148 (2003)
[hep-th/0303109].

\bibitem{AV}
B.~S.~Acharya and C.~Vafa,
{\em On domain walls of ${\cal N} = 1$ supersymmetric Yang--Mills in four dimensions,} hep-th/0103011.

\bibitem{witten3d}
 E.~Witten,
{\em Supersymmetric index of three-dimensional gauge theory,}
in {\sl The Many Faces of the Superworld}, Ed. M. Shifman
(World Scientific, Singapore, 2000), p. 156
[hep-th/9903005].


\bibitem{zarembo}
K.~Zarembo,
Phys.\ Lett.\ B {\bf 462}, 70 (1999) [hep-th/9901106].

\bibitem{zarembo-tseytlin}
A.~A.~Tseytlin and K.~Zarembo,
Phys.\ Lett.\ B {\bf 457}, 77 (1999) [hep-th/9902095].

\bibitem{NSVZ}
V.~A.~Novikov, M.~A.~Shifman, A.~I.~Vainshtein and V.~I.~Zakharov,
Nucl.\ Phys.\ B {\bf 191}, 301 (1981).

\bibitem{divecchiaetal}
P.~Di Vecchia, A.~Liccardo, R.~Marotta and F.~Pezzella,
{\em  On the gauge/gravity correspondence and the open/closed string duality,}
hep-th/0503156.

\bibitem{bergman}
O.~Bergman and M.~R.~Gaberdiel,
Nucl.\ Phys.\ B {\bf 499}, 183 (1997)
[hep-th/9701137].

\bibitem{MN}
J.~M.~Maldacena and C.~Nu\~nez,
Phys.\ Rev.\ Lett.\  {\bf 86}, 588 (2001)
[hep-th/0008001].

\bibitem{KS}
I.~R.~Klebanov and M.~J.~Strassler,
JHEP {\bf 0008}, 052 (2000)
[hep-th/0007191].

\bibitem{polyakov}
A.~M.~Polyakov,
Int.\ J.\ Mod.\ Phys.\ A {\bf 14}, 645 (1999)
[hep-th/9809057].

\bibitem{sagnotti}
A.~Sagnotti,
{\em Some properties of open string theories,}  hep-th/9509080;
Nucl.\ Phys.\ Proc.\ Suppl.\  {\bf 56B}, 332 (1997)
[hep-th/9702093].

\bibitem{MS}
A.~A.~Migdal and M.~A.~Shifman,
Phys.\ Lett.\ B {\bf 114}, 445 (1982).

\bibitem{gra1}
T. Kimura, {\em Divergence of Axial-Vector Current in the Gravitational Field},
Prog.  Theor.  Phys., {\bf 42}, 1191, 1969;
{\em Divergence of Axial-Vector Current in the Gravitational Field. II. Anomalous Commutators and Model Theory}, Prog.  Theor.  Phys., {\bf 44}, 1353, 1970.

\bibitem{gra2}
R.~Delbourgo and A.~Salam,
Phys.\ Lett.\ B {\bf 40}, 381 (1972).

\bibitem{LS}
H.~Leutwyler and M.~A.~Shifman,
Phys.\ Lett.\ B {\bf 221}, 384 (1989).

\end{thebibliography}
\end{document}